\RequirePackage{rotating}
\documentclass[useAMS]{mn2e}
\usepackage{graphicx}
\usepackage{longtable}
\usepackage{rotate}
\usepackage{rotating}
\usepackage{mathtools}
\usepackage{pdflscape}
\usepackage{scalerel}
\usepackage{times}
\usepackage{verbatim}
\newif\ifAMStwofonts

\def\gs{\mathrel{\hbox{\rlap{\hbox{\lower4pt\hbox{$\sim$}}}\hbox{$>$}}}}
\def\ls{\mathrel{\hbox{\rlap{\hbox{\lower4pt\hbox{$\sim$}}}\hbox{$<$}}}}

\def\swift{{\it Swift}}
\def\uvot{{\it UVOT}}
\def\xmm{{\it XMM-Newton}}

\def\nustar{{\it NuSTAR}}

\def\et{{et al.\ }}

\def\mrk335{{Mrk~335}}
\def\izw1{{I~Zw~1}}
\def\rg{{\thinspace r_{\rm g}}}

\def\fvar{{F_{\rm var}}}

\def\A{{\rm\thinspace \AA}}
\def\cm{{\rm\thinspace cm}}
\def\erg{{\rm\thinspace erg}}

\def\ga{{\rm\thinspace gauss}}

\def\keV{{\rm\thinspace keV}}

\def\Msun{\hbox{$\rm\thinspace M_{\odot}$}}

\def\s{{\rm\thinspace s}}

\def\cts{{\rm\thinspace count}}

\def\cps{\hbox{$\cts\s^{-1}\,$}}

\def\ergpscmpspa{\hbox{$\erg\s^{-1}\cm^{-2}\A^{-1}\,$}}

\title[Eleven years of monitoring \mrk335\ ]
      {
Eleven years of monitoring the Seyfert~1 \mrk335\ with \swift: Characterizing the X-ray and UV/optical variability       }

\author[L. C. Gallo \et]
       {L. C. Gallo,$^1$ 
       D. M. Blue,$^{1,2}$
       D. Grupe,$^3$
       S. Komossa,$^4$
 and   D. R. Wilkins$^5$  
        \\ 
$^{1}$ Department of Astronomy and Physics, Saint Mary's University, 923 Robie Street, Halifax, NS, B3H 3C3, Canada \\
$^{2}$ Department of Mathematics, Mount Saint Vincent University, 166 Bedford Hwy, Halifax, NS B3M 2J6, Canada \\
$^{3}$ Space Science Center, Morehead State University, 235 Martindale Drive, Morehead, KY 40351, USA\\
$^{4}$ Max-Planck-Institut f{\"u}r Radioastronomie, Auf dem H\"ugel 69, 53121 Bonn, Germany \\
$^{5}$ Kavli Institute for Particle Astrophysics and Cosmology, Stanford University, Stanford, CA, 94305, U.S.A. \\
}
\date{xxx Accepted. yyy Received. }
\pagerange{\pageref{firstpage}--\pageref{lastpage}}
\pubyear{2017}
\begin{document}
\maketitle
\label{firstpage}

\begin{abstract}
The narrow-line Seyfert 1 galaxy (NLS1) \mrk335\ has been continuously monitored with \swift\ since May 2007 when it fell into a long-lasting, X-ray low-flux interval.  Results from the nearly 11 years of monitoring are presented here.  Structure functions are used to measure the UV-optical and X-ray power spectra.  The X-ray structure function measured between $10-100$~days is consistent with the flat, low-frequency part of the power spectrum measured previously in \mrk335.  The UV-optical structure functions of \mrk335\ are comparable with those of other Seyfert~1 galaxies and of \mrk335\ itself when it was in a normal bright state.  There is no indication that the current X-ray low-flux state is attributed to changes in the accretion disc structure of \mrk335.  The characteristic timescales measured in the structure functions can be attributed to thermal (for the UV) and dynamic (for the optical) timescales in a standard accretion disc.  The high-quality UVW2 ($\sim1800\A$ in the source frame) structure function appears to have two breaks and two different slopes between $10-160$~days. 
Correlations between the X-ray and other bands are not highly significant when considering the entire 11-year light curves, but more significant behaviour is present when considering segments of the light curves.  A correlation between the X-ray and UVW2 in 2014 (Year-8) may be predominately caused by an giant X-ray flare that was interpreted as jet-like emission.  In 2008 (Year-2), possible lags between the UVW2 emission and other UV-optical waveband may be consistent with reprocessing of X-ray or UV emission in the accretion disc.

\end{abstract}



\section{Introduction}
\label{sect:intro}

The narrow-line Seyfert~1 galaxy (NLS1) \mrk335\ ($z=0.025$) has been the subject of intense scrutiny (e.g. Grupe \et 2008; Gallo \et 2013, 2015; Longinotti \et 2008, 2013; Wilkins \et 2015) since it was discovered in a historically low X-ray flux state in 2007 (Grupe \et 2007).  After decades of being one of the brightest AGN in the X-ray sky, \mrk335\ initially faded to about one-thirtieth its typical brightness.  Since that substantial drop, \mrk335\ has never fully recovered to its previous bright state.    Swift monitoring  with the XRT and \uvot\ over the past eleven years find the source to be highly variable in the X-rays, showing persistent variability and occasional intense flaring by factors of ten.  However, on average, the source remains about one-tenth its previous brightness and this low state appears to be the new normal for \mrk335\ (Grupe \et 2012).  It is not clear if the change in behaviour arises from some physical change in the accretion process (e.g. akin to state changes in stellar mass black holes) or simply those produced in a stationary process.

In the optical-to-UV bands the source is one of the most persistently variable radio-quiet AGN changing in brightness by factors of $\sim2$ (e.g. Peterson \et 1998; Grupe \et 2012; Komossa \et 2014). However, during this X-ray dim phase, the average optical/UV brightness has not changed significantly compared to historical values.  Compared to its UV luminosity, \mrk335\ is now in an X-ray weak state (e.g. Gallo 2006).  The substantial dimming of the AGN is only evident in the X-rays.  The origin of the X-ray dimming remains uncertain and explanations have included collapse of the corona perhaps forming a collimated structure (e.g. Gallo \et 2013, 2015; Wilkins \& Gallo 2015; Wilkins \et 2015) or absorption and partial obscuration of the disc and corona (e.g. Grupe \et 2008; Longinotti \et 2013).  While partial covering has difficulty describing the  broadband low- and high-flux intervals in a self-consistent manner (Gallo \et 2015), and the reverberation lags seen in some flux states (Kara \et 2013), there is indication in new \xmm\ observation that obscuration may play a role in the low-flux state (Longinotti \et in prep; Gallo \et in prep).

The long-term, multiwavelength monitoring campaign of \mrk335\ with \swift\ has not been examined to the same degree as the X-ray data (see Grupe \et 2012; Komossa \et 2014).  The UV emission is the dominant component in the spectral energy distribution (SED) of AGN, attributed to the accretion disc that feeds the supermassive black hole.  Changes in the accretion disc structure or flow could explain the weak corona emission we now see in \mrk335.  

The first five years of the monitoring campaign are presented by Grupe \et (2012).  In this work we look at the entire eleven-year monitoring campaign of \mrk335\ that was initiated in 2007 (Grupe \et 2007) with the goal of characterizing the optical/UV emission in the NLS1.  The \swift\ light curves have significant data, but are sampled infrequently because of observing constraints throughout the year.  Therefore, we use appropriate techniques like structure function analysis to examine power spectra and characteristic timescales (e.g. Hughes \et 1992; Collier \& Peterson 2001), and discrete correlation functions (Edelson \et 1988) to investigate correlations and lags.

In the next section the observations and data processing are described.  The light curves and amplitude of the variations are presented in Section~\ref{sect:lc}.  The structure function of each light curve is derived in Section~\ref{sect:sf} and the discrete correlation functions are determined in Section~\ref{sect:dcf}.  Discussion and conclusions follow in Section~\ref{sect:disc} and Section~\ref{sect:conc}, respectively.  Analysis of the spectral energy distribution and spectral variability are saved for future work.

\section{Observations and data reduction}
\label{sect:data}
\begin{table*}
 \centering
 \caption{\swift\ XRT and \uvot\ observations of \mrk335\ since January 2012.  All exposures are given in seconds. Observations prior to 2012 are listed in Grupe et al. (2008) and (2012). The full list of observations is shown online.
 }
  \begin{tabular}{cccccrrrrrrr}
  \hline
  \hline
ObsID & 
Segment & $T_{\rm start}$ (UT) & $T_{\rm end}$ (UT) & MJD & $T_{\rm exp, XRT}$ 
& $T_{\rm exp, V}$ & $T_{\rm exp, B}$ & $T_{\rm exp, U}$ & $T_{\rm exp, W1}$ & $T_{\rm exp, M2}$ & $T_{\rm exp, W2}$ \\
\\
\hline 
35755 & 092 & 55946.819  & 2012-01-20 19:33 & 2012-01-20 19:48 &  919 & --- & --- & --- & --- & --- & 924 \\ 
      & 093 & 55947.828  & 2012-01-21 19:51 & 2012-01-21 19:51 &   25 & --- & --- & --- & --- & --- &  37 \\
      & 094 & 55948.625  & 2012-01-22 14:51 & 2012-01-22 15:08 &  976 & --- & --- & --- & --- & --- & 985 \\
      & 095 & 55949.896  & 2012-01-23 21:22 & 2012-01-23 21:39 &  976 & --- & --- & --- & --- & --- & 973 \\
      & 096 & 55950.965  & 2012-01-24 23:01 & 2012-01-24 23:17 &  954 & --- & --- & --- & --- & --- & 948 \\
\hline
\hline
\label{swift_log}
\end{tabular}
\end{table*}

Mrk 335 has been the target of a continuous survey since May 2007 when it was discovered in an extremely low X-ray flux state (Grupe \et 2007). Here the extremely long (nearly 11 years) \swift\ monitoring of \mrk335\ between 17 May 2007 and 31 December 2017 is reported and the associated optical, UV, and X-ray ($0.3-10\keV$) light curves are presented. The observations were only interrupted during the 3 month period between February to May, when \mrk335\ was in sun constraint with \swift.   The observation log listing all observation between January 2012 and 2018 is provided in Table~\ref{swift_log}.  The log showing data between 2007 and 2011 was shown previously in Grupe \et (2012). 

In order to avoid filter rotations on \swift\ {\it UV-Optical Telescope} (\uvot, Roming et al. 2005), most observations were performed with a 4 or 8 day cadence when the UVW2 ($1928\A$) filter was ``Filter of the Day'', which is a method to reduce filter wheel rotation in order to extend the \uvot\ life time. The X-ray and \uvot\ UVW2 light curves have good sampling over the $\sim11$ years. Observations in the other five \uvot\ filters [V ($5468\A$), B ($4392\A$), U ($3465\A$), UVW1 ($2600\A$), and UVM2 ($2246\A$)] are more sparse. Only at the beginning of the monitoring campaign in 2007 and 2008 was there good coverage in all 6 \uvot\ filters.

All observations with the \swift\ X-ray Telescope 
(XRT, Burrows et al., 2005) were  performed in photon counting mode 
(pc, Hill et al., 2004), except for a short period in 2014, between 17 September and 06 October, when it was observed in Windowed Timing (WT) mode during an X-ray flare event. 
XRT data were reduced with the the task {\sc xrtpipeline} version 0.13.4., 
which is included in the HEASOFT package 16.1. For each observation we extracted source and background spectra and event files using {\sc xselect}. 
Source counts were extracted from a circular region with a
radius of 94$^{''}$ and the background counts were from a source-free, circular
region with a radius 295$^{''}$. We created auxiliary response files (ARF) for
each of these spectra using the {\sc ftool} {\sc xrtmkarf}. 
The most recent response files swxpc0to12s6\_20130101v014.rmf for the pc data and swxwt0to2s6\_20131212v015.rmf for the WT data were adopted.

The \uvot\
data of each observation were co-added in each filter with the \uvot\
task {\sc uvotimsum}. 
Source counts 
 were selected from a circle region with a radius of 5$^{''}$ and the background counts were extracted from 
 a circular region with a radius of $20^{''}$ close the source.   \uvot\ magnitudes and flux densities were measured with the task {\sc  
uvotsource} 
based on the most recent \uvot\ calibration as described in  
Poole \et (2008) and  
Breeveld \et (2010). 
The \uvot\ data were corrected for Galactic reddening ($E_{\rm B-V}=0.035$; Schlegel \et 1998), using
equation (2) in Roming \et (2009) and reddening curves from Cardelli et al. (1989).  No correction has been made for host-galaxy contamination, consequently, the variability in the optical wavebands (e.g. B, V) will be diluted. 

The XRT and \uvot\ measurements, including hardness ratios, for each observation between January 2012 and 2018 are shown in Table~\ref{swift_results}.  The measurements prior to 2012 were listed in Grupe \et (2012).  All uncertainty values in the tables and figures correspond to $1\sigma$ error bars unless specified otherwise.  The timescales are reported in the observed frame.  Given the low redshift of \mrk335, the difference between frames is negligible compared to the timing resolution used in this work. 
\begin{table*}
 \centering
 \caption{\swift\ XRT and \uvot\ measurements of \mrk335\ since January 2012. The XRT values are in units of counts s$^{-1}$. The hardness ratio is defined as $HR = \frac{hard-soft}{hard+soft}$ where soft and hard are the counts in the 0.3-1.0 keV and 1.0-10.0 keV bands, respectively. The reddening corrected \uvot\ magnitudes are in the Vega system.   Measurements prior to 2012 are listed in Grupe et al. (2008) and (2012). The full list of observations is shown online.
 }
  \begin{tabular}{ccccccccc}
  \hline
  \hline
 MJD & XRT CR & XRT HR & V & B & U & UVW1 & UVM2 & UVW2 \\
\\
\hline 
55946.819 &    $0.576\pm0.027$ &   $-0.13\pm0.05$   &    ---         &    ---         &    ---         &    ---         &    ---         & $12.72\pm0.03$ \\ 
55947.828 &     ---           &    ---            &    ---         &    ---         &    ---         &    ---         &    ---         & $12.71\pm0.03$ \\
55948.625 &    $0.367\pm0.021$ &   $-0.05\pm0.05$   &    ---         &    ---         &    ---         &    ---         &    ---         & $12.75\pm0.03$ \\   
55949.896 &    $0.912\pm0.044$ &   $-0.03\pm0.06$   &    ---         &    ---         &    ---         &    ---         &    ---         & $12.78\pm0.03$ \\
55950.965 &    $0.886\pm0.038$ &   $-0.12\pm0.05$   &    ---         &    ---         &    ---         &    ---         &    ---         & $12.82\pm0.03$ \\
55951.098 &    $0.678\pm0.032$ &   $-0.05\pm0.05$   &    ---         &    ---         &    ---         &    ---         &    ---         & $12.82\pm0.03$  \\
\hline
\hline
\label{swift_results}
\end{tabular}
\end{table*}

\section{Light curves and fractional variability}
\label{sect:lc}
\begin{figure*}
	\includegraphics[width=\linewidth]{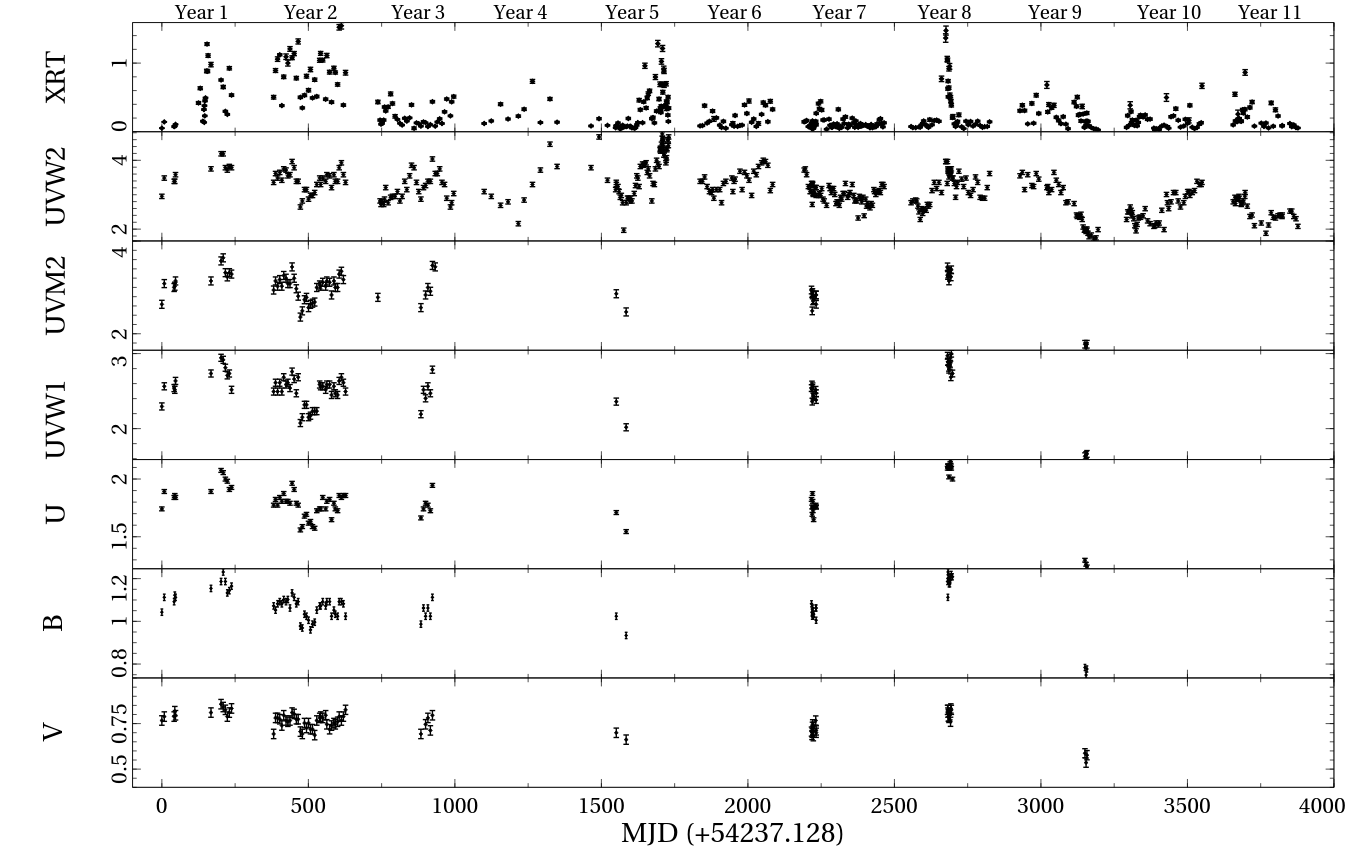}
	\caption{The \swift\ XRT ($0.3-10\keV$) and \uvot\ light curves for \mrk335\ from 17 May 2007 to 31 December 2017.  Data from specific epochs are referred to by ``Year'' as defined in the top row and distinguished by the $\sim3$ month gap where \mrk335\ is in sun constraint with \swift.   The data presented by Grupe \et (2012) are prior to MJD$\sim 54237 + 1750$ (i.e. approximately Year 1 to Year 5).  The XRT light curve is in counts s$^{-1}$.  \uvot\ flux densities are expressed in units of $10^{-14}\ergpscmpspa$.  Error bars in the brightness are present on all data.}
	\label{fig:lc}
\end{figure*}
\begin{figure}
    \begin{center}
        \begin{minipage}{0.9\linewidth}
            \scalebox{1}{\includegraphics[width=\linewidth]{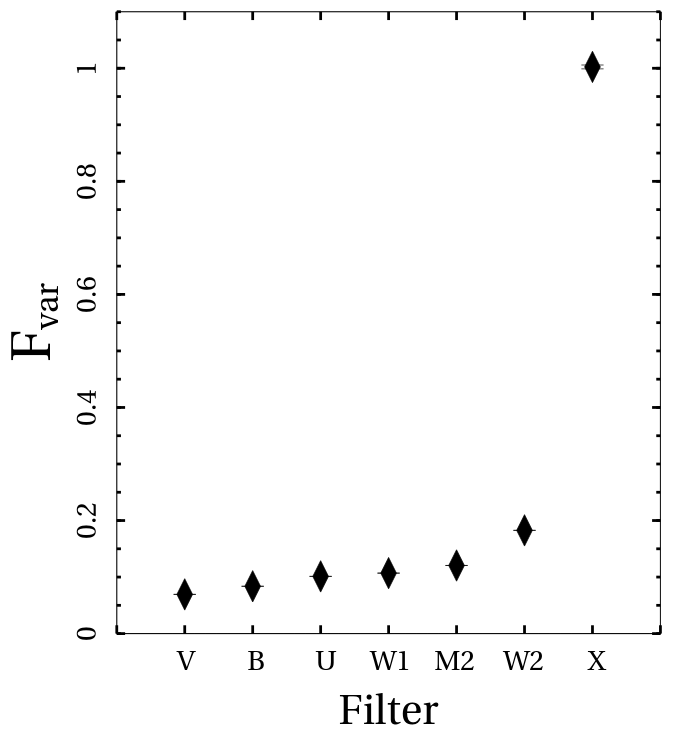}}
        \end{minipage}  \hfill
    \end{center}
    \caption{The fractional variability of \mrk335\ measured in each waveband (shown in increasing energy).  Uncertainties are plotted, but appear smaller than the data points.     }
    \label{fig:fvarT}
\end{figure}

The \swift\ light curves between 2007 May and 2017 December are shown for each waveband in Fig.~\ref{fig:lc}.  In each year (e.g. observing campaign), \mrk335\ is visible to \swift\ for approximately 250 days before it is in sun constraint.  The cadence of the observations depends on the filter used.  The X-rays ($0.5-10\keV$) and UVW2 are the preferred combination and the typical sampling rate is between $4-8$~days during an observing year.  However, increased monitoring is requested at times and some parts of the light curve may even be sampled daily.  The other optical/UV filters are used less frequently.  Data obtained since January 2012 are presented in Table~\ref{swift_log} and \ref{swift_results} (full tables are available online).  Data obtained prior to 2012 are available from Grupe \et (2008, 2012).

The light curves exhibit significant variability in each waveband with the largest fluctuations apparent in the X-ray.  The minimum-to-maximum X-ray count rate varies by a factor of $\sim50$ over the eleven years from $\sim0.03$ to $\sim1.54\cps$. 

To estimate the amplitude of the variations in each waveband, the fractional variability ($\fvar$) and uncertainties are calculated following Edelson \et (2002).  The $\fvar$ for the entire 11-year data set shows the amplitude of the variations increase with energy (Fig~\ref{fig:fvarT}).  The X-ray band is significantly more variable than the UV and optical bands.   The $\fvar$ decreases from $\sim20$ per cent in the UVW2 filter to $\sim7$ per cent in the V band.   This diminished variability at longer wavelengths likely results from a combination of increased host-galaxy contamination and decreased intrinsic variability.  Since the amplitude of the variations in the UV and optical are small compared to X-rays (Fig.~\ref{fig:fvarT}), the signals (e.g. correlations and delays) searched for in the analysis are expected to be weak. 

\begin{figure*}
    \includegraphics[width=\linewidth]{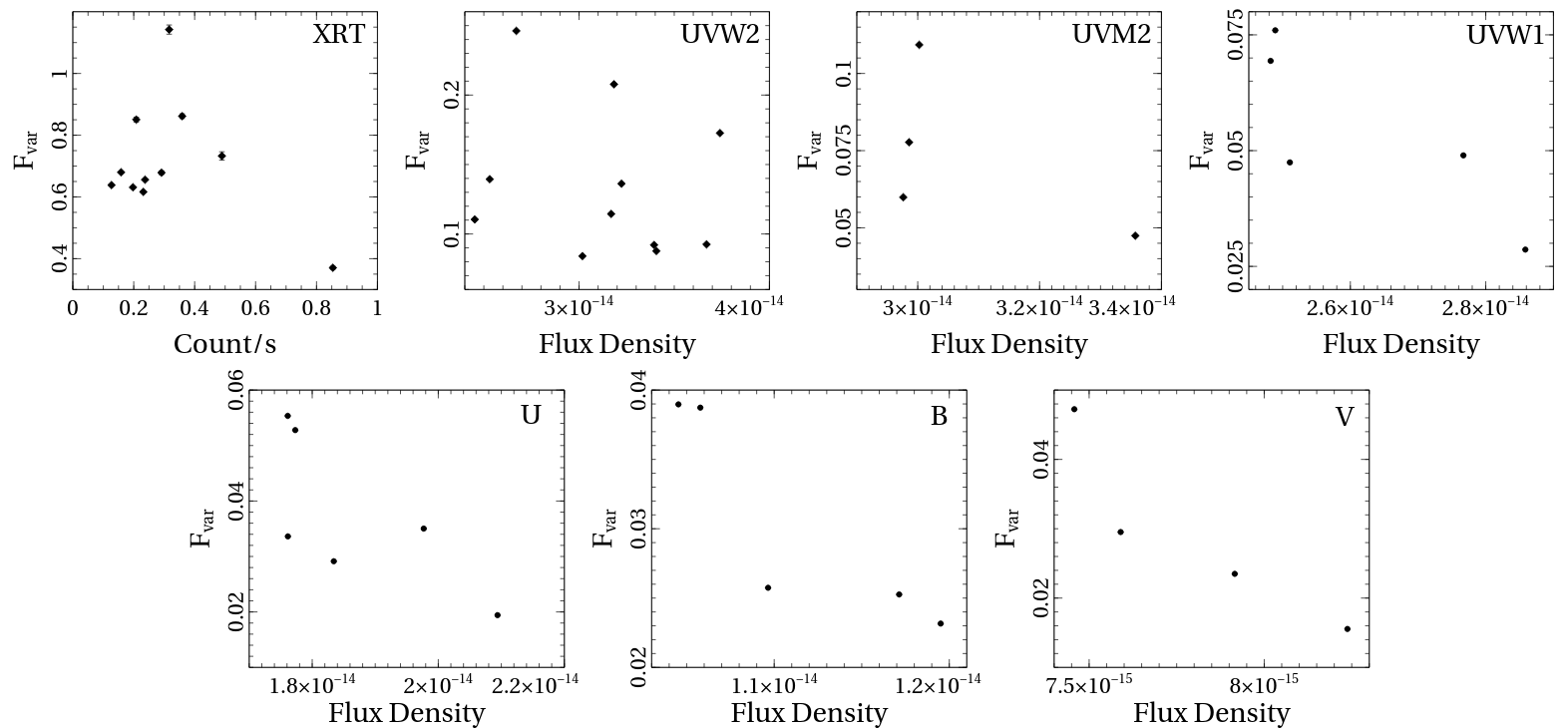}
    \caption{The fractional variability of \mrk335\ measured in each waveband plotted against the mean flux during that observing year. Each data point corresponds to a single campaign (year) in which at least six observations were completed. Uncertainties are plotted, but appear smaller than the data points.  \uvot\ flux densities are expressed in units of $\ergpscmpspa$.
    }
    \label{fig:fvar}
\end{figure*}

The fractional variability is calculated for each waveband in each observing year, for which at least six observations were obtained.  All wavebands exhibit a common trend in which the $\fvar$ diminishes with increasing flux (Fig~\ref{fig:fvar}), which is indicative of a stationary process.  However, there is significant scatter in how $\fvar$ changes with flux and time.  This does not necessarily indicate changes in the underlying physical processes generating the energy.  The measured fractional variability does depend on the sampling in a given filter,  which does vary from year-to-year.  This likely contributes to the significant scatter present in the $\fvar$-flux trend in all filters.  A stationary process dominated by red noise will exhibit differences in the mean and variance with time (e.g. Vaughan \et 2003). 

\section{Structure Function Analysis}
\label{sect:sf}

\subsection{The Structure Function}

The power spectral density (PSD; power spectrum) shows the distribution of variability power in a light curve as a function of frequency.  For active galactic nuclei, the power spectrum follows a power law shape ($P(f) \propto f^{-\alpha}$, where $P$ is the power and $f$ is the temporal frequency), with index $\alpha\sim2$ (i.e. red noise) (e.g. McHardy \et 2004; Ar\'evalo \et 2008; Breedt \et 2010).  Toward lower frequencies (longer timescales), the spectrum will flatten ($\alpha\sim1$).  The break or characteristic frequency ($f_b$) is attributed to some underlying physical process and scales  with the black hole mass and accretion rate (e.g. McHardy \et 2006).  There is also indication it may be linked to the line-of-sight absorption in the source (Gonz\'alez-Mart\'in 2018).
Work on stellar mass black hole binaries shows that the variability in the shape of the power spectrum can be used to reveal underlying changes in the behaviour of the accretion system that are not easily disclosed with a fractional variability analysis. 

The \mrk335\ data set is rich, but it is inhomogeneous and the sampling is uneven over the course of a year.  Therefore, using Fourier transform techniques, like the PSD, are possible, but can be problematic.  
For this work, the distribution of power in the light curves is examined using a structure function analysis (SF; e.g. Simonetti \et 1985; Hughes \et 1992; Di Clemente \et 1996).   

The method of structure function analysis to study variability of ensembles and individual AGN is widely used (e.g. Simonetti \et 1985; Hughes \et 1992; Di Clemente \et 1996; Vanden Berk \et 2004; Bauer \et 2009; Vagnetti \et 2016; Middei \et 2017).  Collier \& Peterson (2001, hereafter CP01) apply the method to study the UV/optical variability in a sample of Seyfert galaxies including \mrk335.

The SF can be used in the same manner as a PSD to evaluate the distribution of power in a time series.  The primary difference is the structure function operates in the time domain and can be used effectively to examine irregularly sampled light curves like those in this work.  Specifically, the SF is a function of time differences ($\tau = t_j - t_i$) between pairs of points, $i$ and $j$ (where $j > i$), in a series with $N(\tau)$ pairs.  The definition used by CP01 is adopted here to ease comparison to that work. That is. 
\begin{equation}
         SF\left(\tau\right)=\frac{1}{N\left(\tau\right)}\sum_{i<j}\left[f\left(t_i\right)-f\left(t_i+\tau\right)\right]^{2}
\label{eq:sf}
\end{equation}

Between some minimum and maximum timescale where the variations are correlated, the SF will take on a power law shape.  Like the PSD, the slope of the spectrum encodes information of the accretion process and for a stationary light curve the PSD slope is related to the slope of the SF ($\beta$) by $\alpha=\beta+1$ (Hufnagel \& Bregmann 1992; Kawaguchi \et 1998).   However, Paltani (1999) note this is this is only valid for an infinite time series.

At long timescales the SF  will begin to flatten at approximately two-times the time series variance ($2\sigma^{2}$) (Hughes \et 1992).  The turnover occurs at some characteristic timescale ($\tau_{char}$) that is related to the break frequency in the PSD.  The power law also flattens at short timescales where the signal is approximately two-times the noise variance ($2\sigma^{2}_{n}$) (Hughes \et 1992).  However, the flattening could also be an artifact of a finite, non-stationary light curve.

CP01 test if the observed flattening in the SF could be attributed to physical processes and if the power law slope ($\beta$) could be related to the PSD slope ($\alpha$).  CP01 find that a power law slope is expected at times $\ls 0.3T$, where $T$ is the duration of the light curve.  Any flattening occurring at timescales less than $0.3T$ may represent flattening from intrinsic AGN processes.  Likewise, the power law slope can be used to predict the PSD spectral index as noted above, but sampling and binning of the structure function may invalidate the relation.  CP01 estimate the systematic uncertainty to be at the 10 per cent level so the simple relation between $\alpha$ and $\beta$ can be used as an approximation.

The method employed in this work is the same as that of CP01.  The binned SF is calculated using Equation~\ref{eq:sf}.  Binning is achieved such that the midpoint of bin $i$ is at $\tau_i = (i-\frac{1}{2})\delta$, where $\delta$ is the resolution of the structure function taken to be the median temporal sampling of the light curve with duration $T$.  The SF is only calculated for bins with more than six light curve pairs.  The number of bins for each SF is $\frac{T}{\delta +1}$.  Two-times the mean noise variance is subtracted from each SF bin to remove the effects of the measurement uncertainties.  This will also result in the flat plateau at the short timescales to be minimized in the SF.  Finally, the SF is then normalised to the light curve variance to ease comparison to CP01.  Again following CP01, the statistical uncertainties in the SF are defined as $\frac{\sigma_i}{\sqrt{N_i/2}}$, where $N_i$ is the number of pairs in bin $i$ and $\sigma_i$ is the root mean square deviations about the mean SF value in that bin.

\subsection{Optical and UV Structure Functions}

\begin{figure*}
    \includegraphics[width=\linewidth]{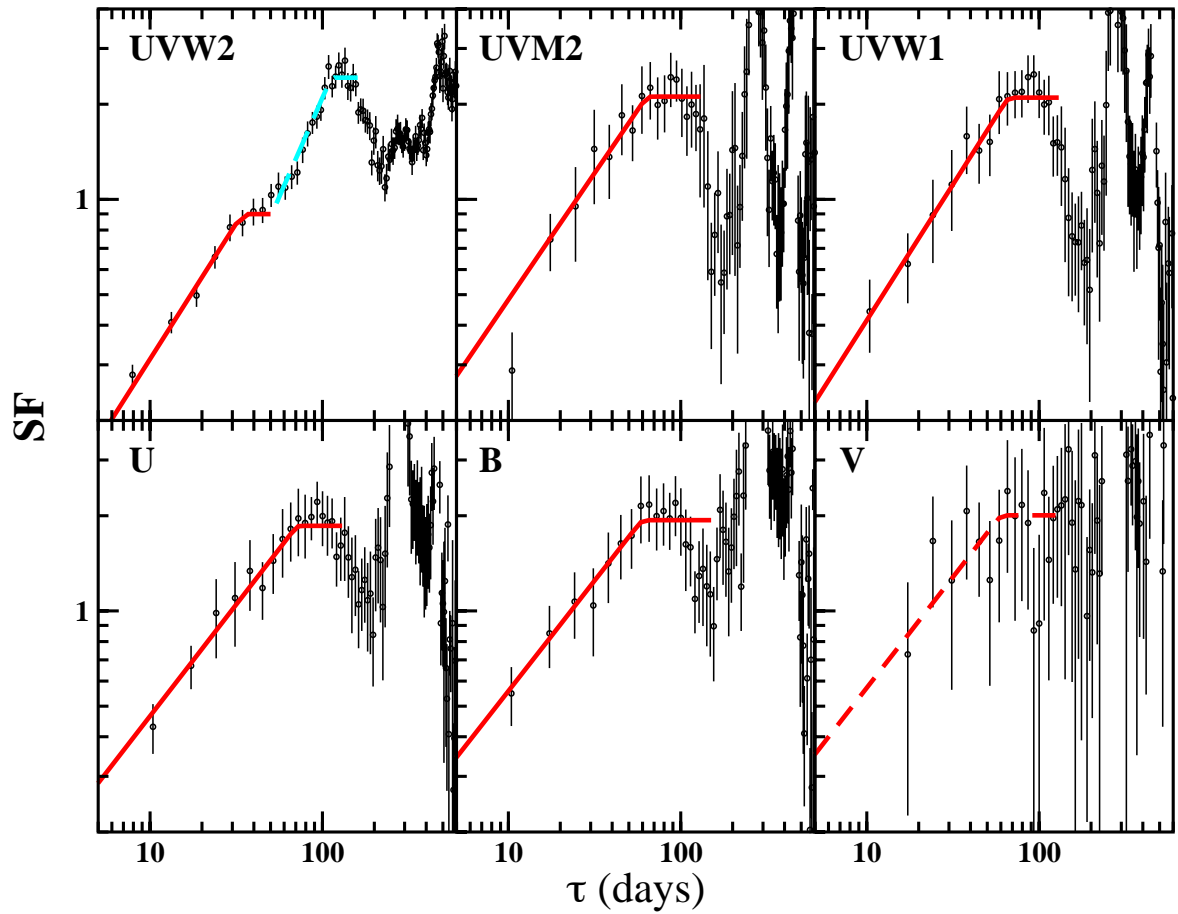}
    \caption{Broken power laws fitted to the structure functions of the \swift\ \uvot\ observations of \mrk335.  The typical SF show a power law slope with increasing $\tau$ before flattening.  The flattening continues briefly before the onset of the oscillations in the noise-dominated regime of the SF.  For clarity, the data are only shown up to $\tau=600$~days.   Two fits of the structure function are done for the UVW2 data. (a) corresponds to  fitting the SF between $10-50$~days (red solid model in top-left panel) and (b) corresponds to fitting the data between $50-160$ days (dashed cyan model in top-left panel) to demonstrate the possibility of two distinct slopes and characteristic timescales in the UVW2 SF.  A fit to the V-band data could not be well constrained.  The model shown (dashed red curve) correspond to the B-band SF fit.}
    \label{fig:uvotsf}
\end{figure*}
\begin{table*}
    \begin{center}
        \caption{The \swift\ \uvot\ and XRT structure function parameters measured for \mrk335. The waveband is given in column (1) followed by total observation length $T$ and structure function resolution $\delta$ in columns (2) and (3), respectively.  Column (4) and (5) are the minimum  ($\tau_{min}$) and maximum ($\tau_{max}$) time lag over which the broken power law model is fit.  The measured parameters of the structure function are the power law slope ($\beta$ ; column 6) and the break in the power law (i.e. characteristic timescale) ($\tau_{char}$; column 7).  where the break occurs in the SF power law.  The  value of the power law slope corresponding to the PSD ($\alpha$) is listed in Column (8) assuming $\alpha=\beta +1$.  Two fits of the structure function are done for the UVW2 data. (a) corresponds to  fitting the SF between $10-50$~days and (b) corresponds to fitting the data between $50-160$ days. *A fit to the V-band data could not be well constrained.  The values shown correspond to the B-band SF fit.
        }
\begin{tabular}{cccccccc}
            \hline
            (1) & (2) & (3) & (4) & (5) & (6) & (7) & (8) \\
            Waveband & $T$ & $\delta$ & $\tau_{min}$ & $\tau_{max}$ & $\beta$ & $\tau_{char}$ & $\alpha$ \\
                 & (days)   & (days) & (days) & (days) &  &(days) & (PSD) \\
            \hline
            \hline
            X-ray              & 3524.232 & 5.3 & 10  & 100        & $0.14\pm0.05$                & $--$                         & $1.14$ \\
            UVW2$^{a}$ & 3524.232 & 5.3 & 10     & 50     & $0.86^{+0.14}_{-0.13}$    & $34^{+11}_{-5}$      & $1.86$ \\
            UVW2$^{b}$ & 3524.232 & 5.3 & 50     & 160       & $1.25^{+0.20}_{-0.18}$    & $112^{+9}_{-7}$     & $2.25$ \\
            UVM2            & 3157.063 & 7.0 & 10   & 130    & $0.80^{+0.29}_{-0.50}$     & $63^{+32}_{-21}$   & $1.80$ \\
            UVW1           & 3157.063 & 6.9 & 10   & 130     & $0.86^{+0.27}_{-0.21}$    & $66^{+21}_{-14}$    & $1.87$ \\
            U                   & 3157.063 & 6.9 & 10   & 130     &$0.71^{+0.28}_{-0.21}$     & $71^{+27}_{-18}$    & $1.71$ \\
            B                   & 3157.063 & 6.9 & 10   & 130     & $0.71^{+0.30}_{-0.51}$     & $59^{+33}_{-19}$    & $1.71$ \\
            V$^{*}$          & 3157.063 & 6.9 & 10   & 130     & $0.71$                              & $59$                        & $1.71$ \\
            \hline
            \label{tab:sf}
        \end{tabular}
    \end{center}
\end{table*}

The structure functions for the approximately eleven-year light curves of \mrk335\ in the \uvot\ bands are shown in Fig.~\ref{fig:uvotsf}.  All the SFs, except for the V-band, exhibit the power law slope with the flattening at long timescales before oscillations arise at timescales when the SFs are poorly defined.   The flattening at short timescales is less obvious as the measurement uncertainties were subtracted when building the SFs  However,  there may be some indication of flattening in some SF at $\tau<10$~days.  

The power law shape is well defined in all SFs except for the V-band.  Between $\tau_{min}$ (taken to be 10~days for all SFs) and $\tau_{max}$ ($\gs100$~days), each structure function is fitted with a broken power law where the break is the transition between the sloped and flat part of the function.  The break timescale is defined as the characteristic timescale ($\tau_{char}$).  The slope above $\tau_{char}$ is fixed at 0 to ease the fitting process.  Determining the break timescale is rather subjective because the flat part of the SF is of short duration before the function becomes noise dominated (i.e. the oscillations set in).  Therefore, the values of $\tau_{char}$ should be considered with caution.   The power law and characteristic timescale are always measured over a range that is $< 0.3T$ thus the results should be robust to biases discussed in Section~\ref{sect:sf}.  The characteristics of each SF and the fit results are shown in Table~\ref{tab:sf}.  In the final column of Table~\ref{tab:sf}, the corresponding PSD slope is listed assuming $\alpha=\beta+1$ as discussed above.

The V-band light curve is crudely sampled and exhibits the weakest variability (i.e. lowest $\fvar$), thus the poorly defined SF is not surprising.  The broken power law could not be fitted to the data.  In Fig.~\ref{fig:uvotsf}, the V-band SF is shown with the B-band model overplotted.  Since the V and B are adjacent bandpasses, the SFs are expected to be similar as appears to be the case (Fig.~\ref{fig:uvotsf}).  

The values of the SF slope are $\sim0.71$ for the optical bands (U, B, and V) and $\sim0.80$ for the UV bands (UVW2 between [$\tau = 10-50$ days], UVM2, UVW1) though all measurements agree within uncertainties.  This is suggestive of a similarly physical process driving the variations in the optical and UV bands.  The corresponding PSD slopes are $\sim 1.7-1.8$, which are typical of most Seyfert~1 galaxies, though there are exceptions (e.g. Mushotzky \et 2011).  The characteristic timescale is about $60$~days though there is a large range on the uncertainty on each individual measurement.  

Interestingly, the high-quality data in the UVW2 light curve, reveals complexity in its SF (Fig.~\ref{fig:uvotsf}, top left).  There is evidence of multiple breaks and different slopes in the power spectrum.  Fitting a broken power law between $10-50$~days reveals a slope of $\beta\approx0.86$ and a break at about $35$~days.  Between $50-160$~days, the slope is steeper ($\beta\approx 1.25$) and a second break is evident at $\tau_{char}\approx112$ days.  Multiple breaks were also seen in the optical SF of NGC~5506 (CP01).


\subsection{X-ray structure function}
\begin{figure}
    \begin{center}
        \begin{minipage}{0.99\linewidth}
            \scalebox{1}{\includegraphics[width=\linewidth]{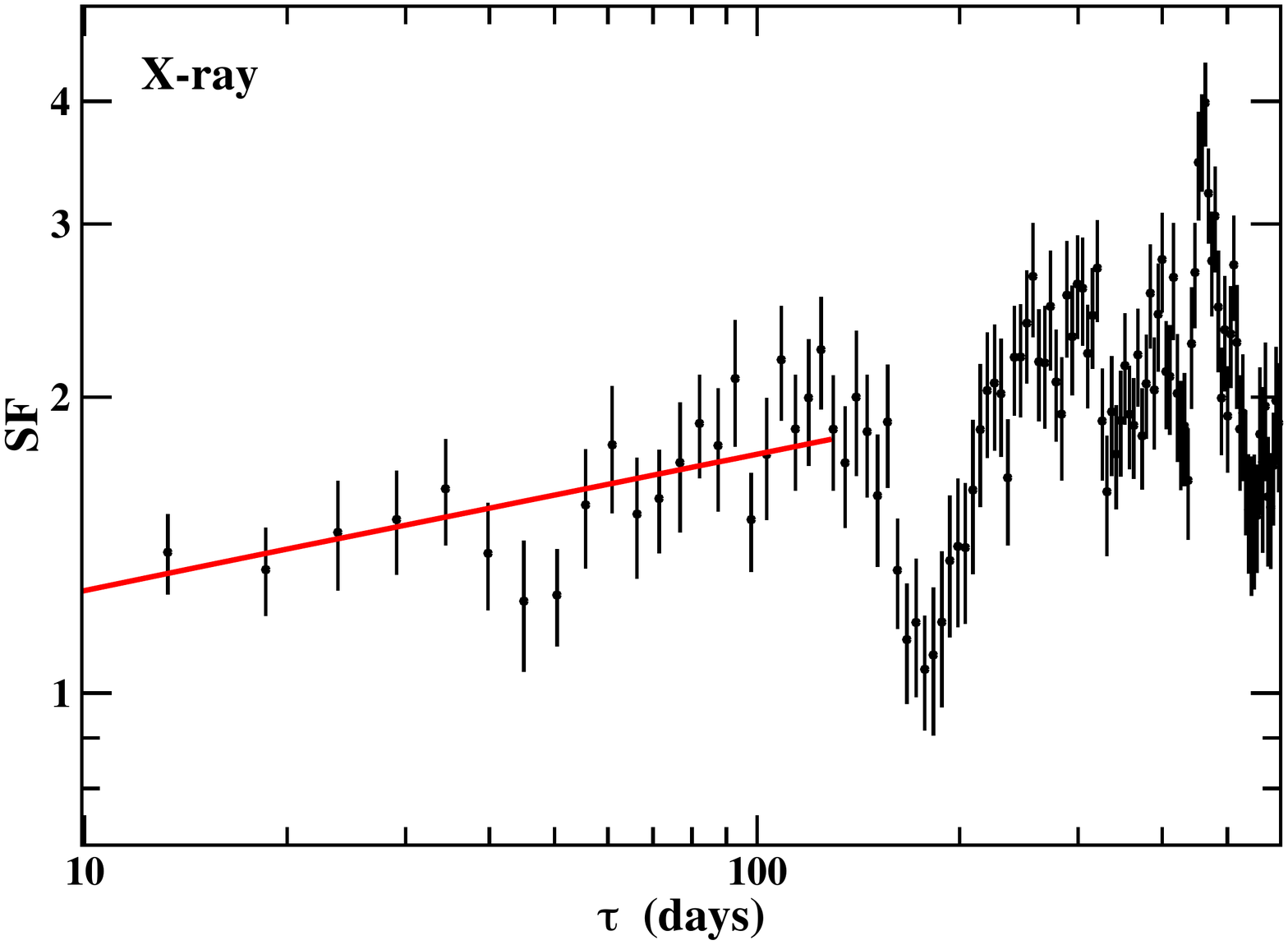}}
        \end{minipage}  \hfill
    \end{center}
    \caption{The power law fitted to the X-ray structure function between $10-100$~days {\bf is consistent with a PSD slope of $\sim1$.}
    }
    \label{fig:xsf}
\end{figure}

The X-ray structure function is different than than the UV/optical ones (Fig.~\ref{fig:xsf}).  The function is significantly flatter ($\beta=0.14\pm0.05$) and there is no measurable evidence of a break in the power law before oscillations set in at $\sim120$ days.   It may even be possible the oscillations start sooner seeing the two SF bins below $100$~days that are outliers to the fit.  

The SF is examined between $10-100$~days and these are sufficiently low frequencies that the break likely occurs at much shorter timescales (higher frequencies).  Ar\'evalo \et (2008) measure a low frequency break in \mrk335\ at $\sim1.6\times10^{-4}$~Hz ($\sim0.07$ days) so it could be we are not examining the steep part of the power spectrum in the X-rays.  The low-frequency slope in the PSD measured by Ar\'evalo \et (2008) is $\sim 1.1$, which agrees with our SF measurement of $\beta$ that would correspond to $\alpha\approx1.14$.  While the \swift\ X-ray sampling is not probing the linear part of the SF, the low-frequency behaviour is consistent with a PSD slope of $\sim1$. Analysis of the X-ray power spectrum would improved from examining the long \xmm\ observations with more frequent sampling on the order of hours and days.


\section{Discrete Correlation Function}
\label{sect:dcf}

The correlations between light curves is examined over the entire $\sim11$-year period as well as within each of the yearly observing campaigns.  Year 2 (Fig.~\ref{fig:lc}) is fruitful to examine correlations among all the \uvot\ since the light curves were well sampled in all wavebands that epoch. 

Given the irregular sampling in the time series, the Discrete Correlation Function (DCF; Edelson \& Krolik 1988) is employed to investigate correlations between light curves.\footnote{The code developed by Robertson \et (2015) is used to calculate the DCF and errors following Edelson \& Krolik (1988).}.    Monte Carlo confidence intervals are determined for each DCF.  A PSD slope of $\alpha = 1.2$ (see Section~\ref{sect:sf}) is adopted to generate $10^5$ random X-ray light curve (Timmer \& K\"onig 1995) with similar binning and cadence to the true X-ray data.  Gaps are inserted in the light curves to simulate the periods of sun constraint in the real data.  Artificial X-ray light curves are compared with the true optical-UV light curves and the DCF coefficients are determined at each lag interval.  A cumulative distribution of the DCF coefficients is constructed to determine the 95 and 99.9 per cent coefficients that are show in the plots.  The process is similar for generating UVW2 light curves that are compared to the other \uvot\ light curves in Year 2.

\subsection{Correlations with X-rays over the eleven years}

\begin{figure*}
    \begin{center}
        \begin{minipage}{0.99\linewidth}
            \scalebox{1}{\includegraphics[width=\linewidth]{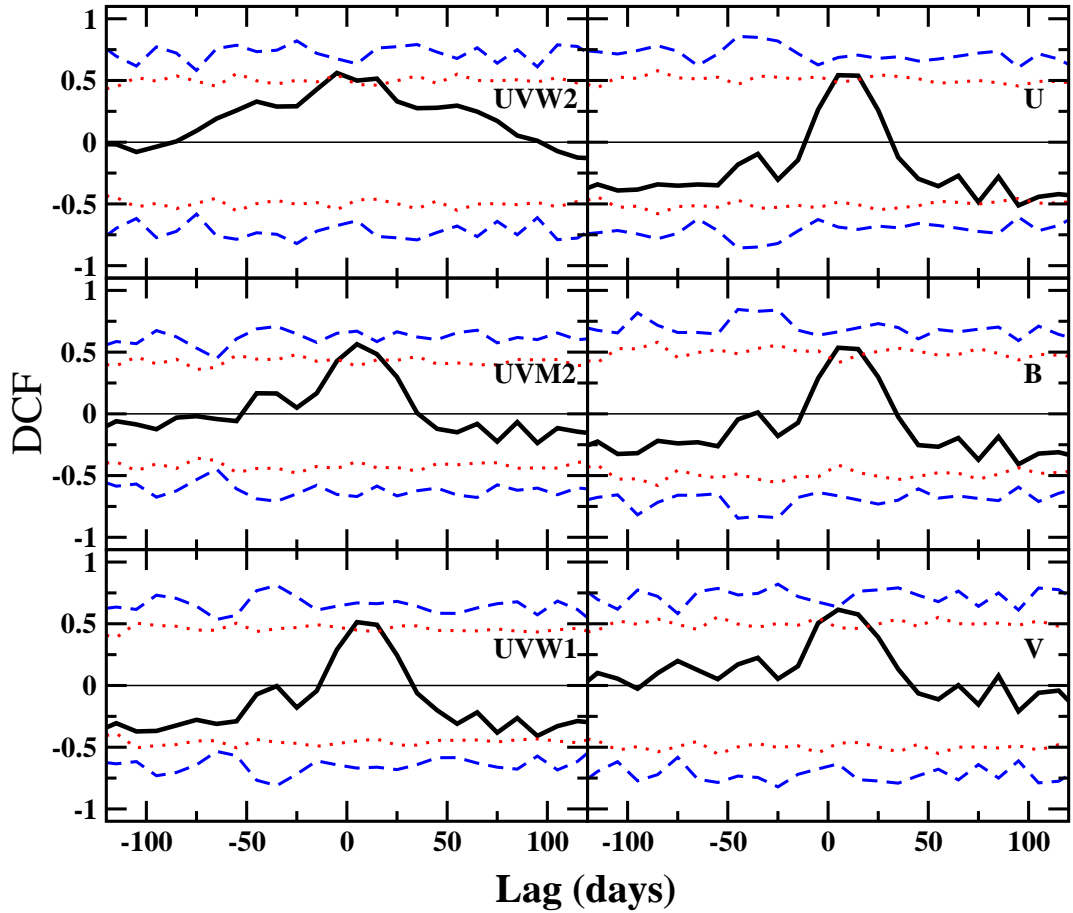}}
        \end{minipage}  \hfill
    \end{center}
    \caption{The \swift\ DCFs between the X-rays and \uvot\ filters over the 11 years of monitoring \mrk335. Correlations are shown between $\pm120$~days and positive lags imply the X-rays are leading the second band (shown in each panel).  The horizontal line marks zero correlation while the dotted-red and dashed-blue curves denote 95 and 99.9 per cent confidence contours, respectively.
    }
    \label{fig:335dcfil}
\end{figure*}
The DCFs between the X-ray light curve and each UV-optical waveband for the entire 11-year monitoring campaign are shown in Fig.~\ref{fig:335dcfil}.  The peak correlation is significant at $\sim95$ per cent confidence for all wavebands, but never reaches the $99.9$ per cent confidence contour.  

The X-ray and UVW2 light curves exhibit a DCF peak at approximately zero lag.  However, for all other wavebands, the DCFs exhibit an asymmetry toward a positive lag.  The peaks are at $5\pm5$~days, which corresponds to one resolution element and are formally consistent with zero.  

The similarity suggests that a short lag does exist, such that the X-rays lead the UV-optical bands, but light curves with higher time-resolution are required, as implied by Ar\'evalo \et (2008).   This may be indication of the reprocessing of X-rays in the accretion disc (e.g. Cameron \et 2012; Shappee \et 2014; McHardy \et 2014).

\subsection{A comparison between the X-ray and UVW2 light curves in each year}
\begin{figure}
    \begin{center}
        \begin{minipage}{0.99\linewidth}
            \scalebox{1}{\includegraphics[width=\linewidth]{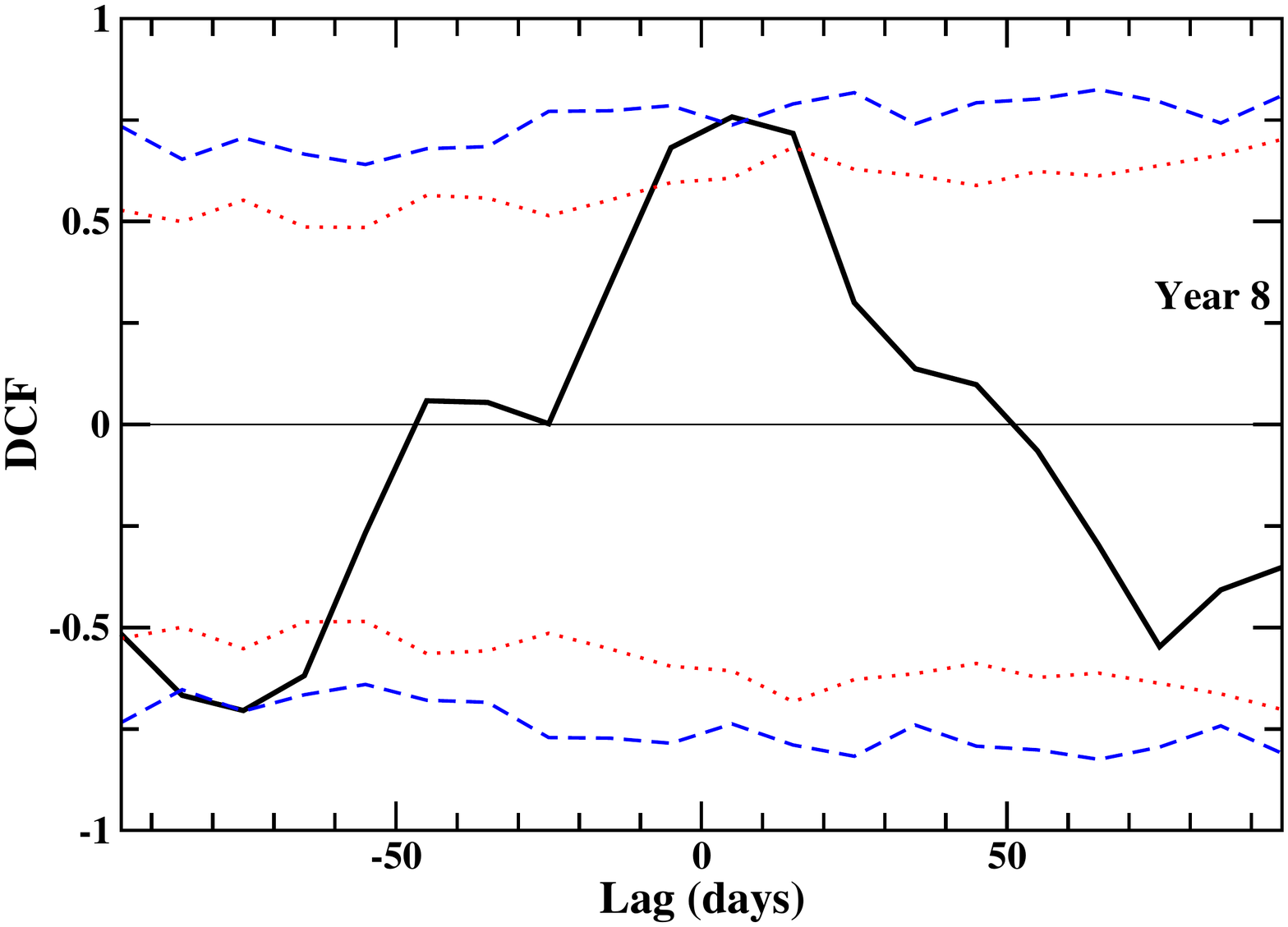}}
        \end{minipage}  \hfill
    \end{center}
    \caption{The \swift\ DCF between the X-rays and UVW2 filter during Year 8 of monitoring \mrk335. Year~8 exhibits the strongest single-year correlation between the X-ray and UVW2. Correlations are shown between $\pm95$~days and a positive lag implies the X-rays are leading the UVW2.  The horizontal line marks zero correlation while the and dotted-red and dashed-blue curves denote  95~ and 99.9 per cent confidence contours, respectively.  
    }
    \label{fig:Y8}
\end{figure}

The UVW2 light curve is the best sampled of all the optical-UV wavebands.  Given the good simultaneous sampling between the X-rays and UVW2 light curves, the DCF for this pair can be calculated for each individual year.  The errors on the DCF are rather large for Years 1 and 4, when there are limited data, so little can be said of those epochs.  For the remaining epochs, the correlations during each individual year do not exceed the $95$ per cent confidence level except for Year 8 (Fig.~\ref{fig:Y8}).  

Like many of the DCFs in the eleven-year average (Fig.~\ref{fig:335dcfil}), the DCF in Year 8 exhibits a peak at a positive lag between $0-10$ days, but only in this case is the significance of the correlation at the $\sim99.9$ per cent confidence level (Fig.~\ref{fig:Y8}).  Similar to the  eleven-year average, the lag is consistent with the X-rays leading the UV band. 

Notably, Year 8 (2014) corresponds to when \mrk335\ displayed a giant X-ray flare accompanied by more modest brightening in the UVW2 filter (Wilkins \et 2015).  It could be the DCF in Year 8 is dominated by this one extreme event making the correlation more significant.

\subsection{Correlations between all light curves in Year 2}

The monitoring of \mrk335\ during the second year was the most complete in that it included measurements in nearly all filters at each epoch.   This makes it possible to examine for correlations between light curves in the UVW2 and all other UV-optical filters, as well as comparing with the X-rays.  

\begin{figure*}
   \centering
   \advance\leftskip-0.cm
   {\scalebox{0.36}{\includegraphics[trim= 1.cm 0cm 1.5cm 0cm, clip=true]{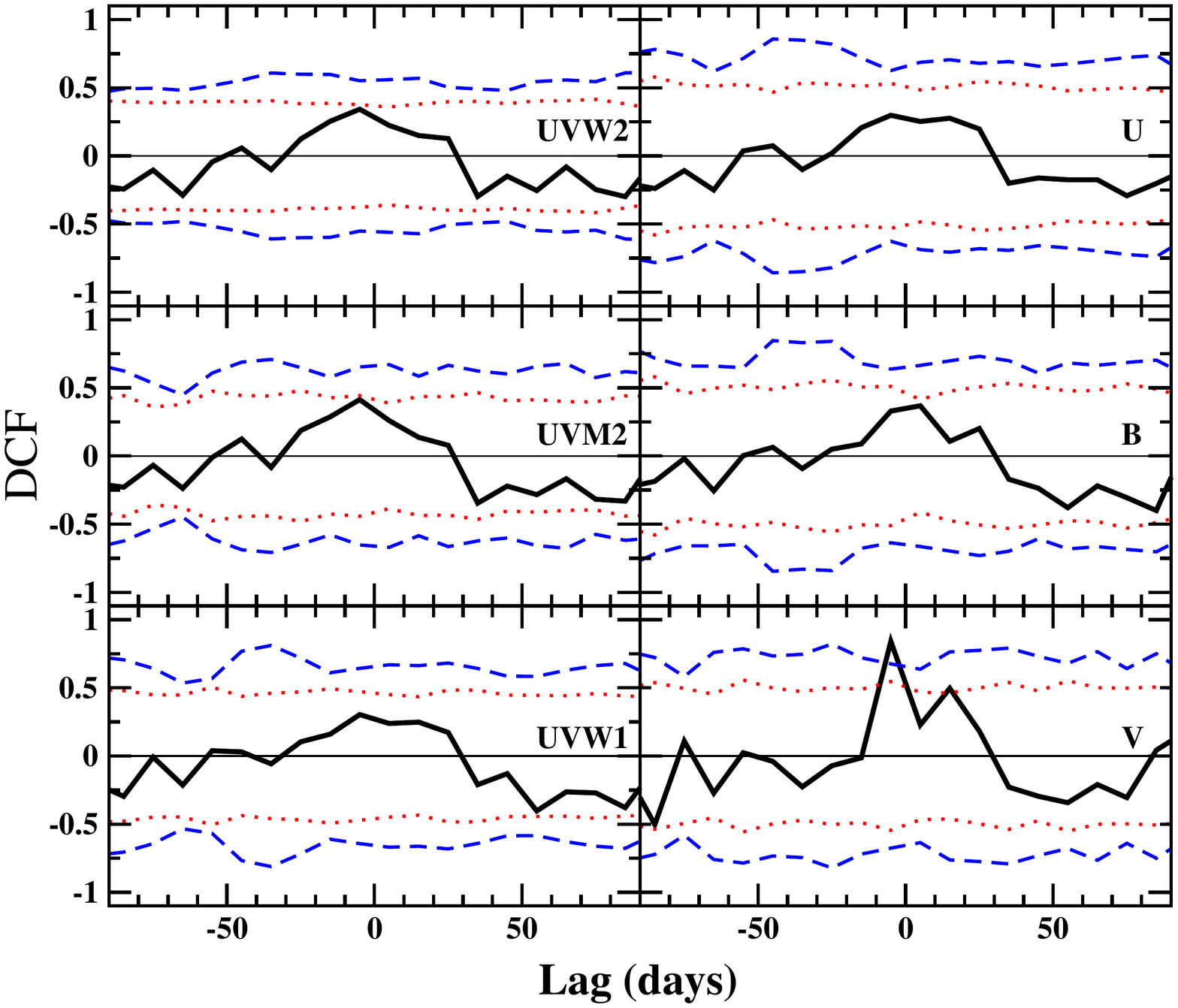}}}      
   {\scalebox{0.36}{\includegraphics[trim= 1.0cm 0cm 3.5cm 0cm, clip=true]{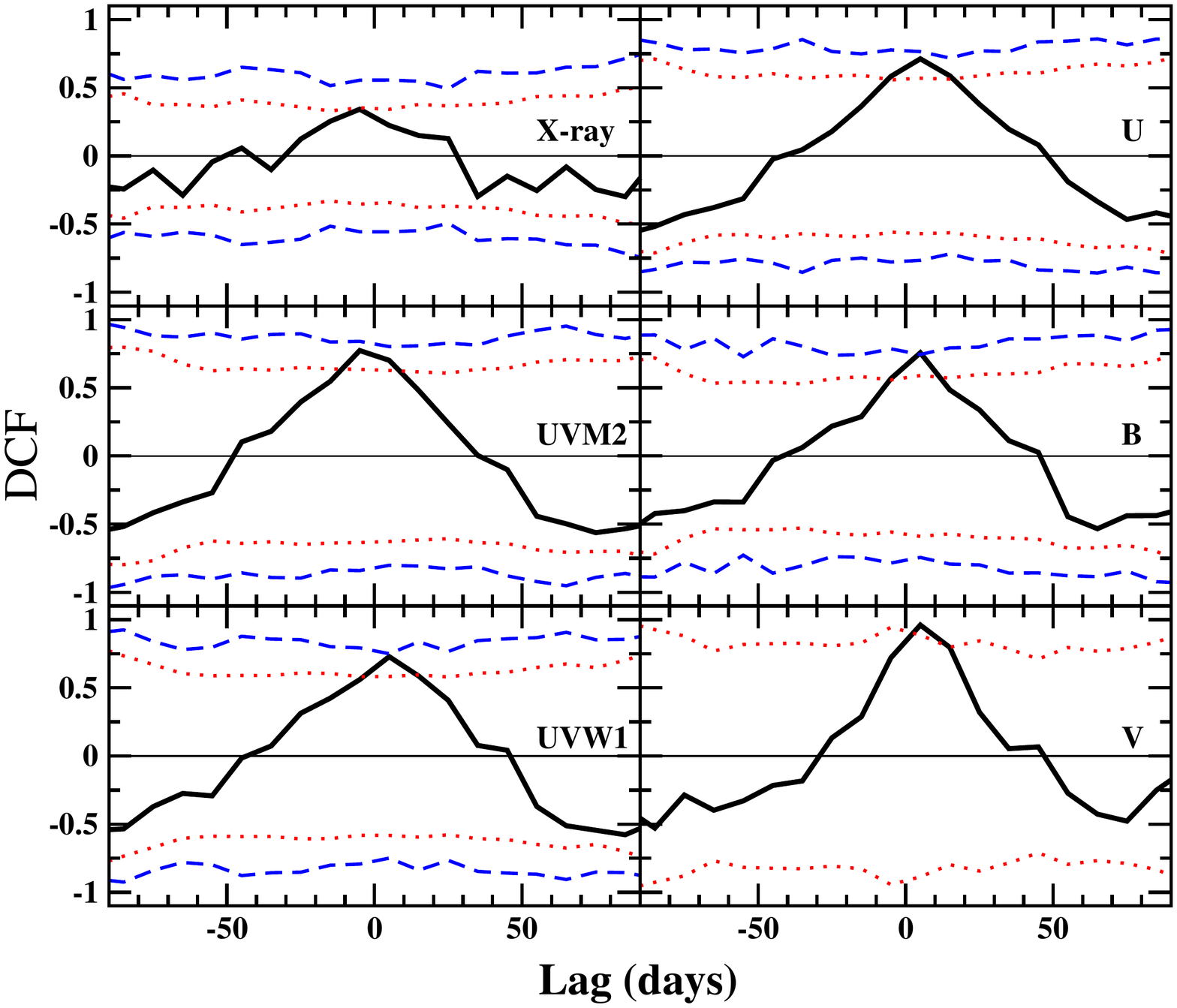}}}       
   \caption{Left panel: The \swift\ DCFs of \mrk335\ for the X-rays and the various \uvot\ filters during Year 2 only. No significant correlations are revealed and the sharp peak in the V-band DCF is likely a spurious event.
Right panel:   The \swift\ DCFs between the UVW2 and all other \uvot\ filters (and X-ray) during Year 2 only.  The top-left panel (UVW2 and X-ray) is the same as in the left panel.  Correlations between the UVW2 and UVW1, U, and B wavebands are significant at $>95$ per cent confidence.  In addition, the peaks in these DCFs are offset to a positive lag, which implies the UVW2 variations are leading.  Note that the UVM2 shows a peak at zero, which is expected since the wavebands overlap in energy.   The correlations with the V band are not significant.  The horizontal line marks zero correlation while the dotted-red and dashed-blue curves denote 95 and 99.9 per cent confidence contours, respectively. }
\label{fig:dcfy2}
\end{figure*}

In Fig.~\ref{fig:dcfy2}, the DCF is shown for each UV-optical light curve against the X-rays (left panel) and against the UVW2 (right panel) light curves during Year 2.  The DCFs comparing the X-rays variations to the UV-optical light curves show no significant correlations.  The strongest peak is seen at $\sim95$ per cent confidence level  in the UVW2 and UVW1 DCFs (left panel).  There is, what appears to be a single significant data point in the V-band DCF that peaks at negative lag.  Given the low-amplitude variability in this band, it seems more likely the single data point is a spurious event rather than a significant lag.  Indeed, if the centroid of the DCF were used to measure the lag rather than the peak (e.g. Welsh 1999), the delay would not be important.

The DCFs comparing the UVW2 light curve with some other UV-optical wavebands (i.e. UVW1, UVM2, U, and B) yield more significant correlations at $>95$ per cent confidence.  The correlation between UVW2 and V are not significant.  The correlation  between UVW2 and UVM2 is consistent with zero lag.  This is expected since both filters examine overlapping parts of the spectral energy distribution.  

The peak of the DCFs between UVW2 and  UVW1, U, and B, are all offset by $\sim 5$~days, indicating that changes in shorter wavelength band (UVW2) are leading variations in the other wavebands.  As with the eleven-year average DCFs, the lags corresponds to one resolution element and are formally consistent with zero.  However, seeing the same offset in all the wavebands suggest that that interband lags may be occurring on timescales shorter than 5~days.  Again, more frequent monitoring with shorter cadence may better constrain these lag estimates.  There are segments in this \swift\ light curve that are sampled with higher frequency, but these are not sufficient to carry out the analysis. 

\section{Discussion} 
\label{sect:disc}

\begin{figure*}
	\scalebox{0.7}{\includegraphics[trim= 1cm 1cm 1.5cm 1cm, clip=true,width=\linewidth, angle=-90]{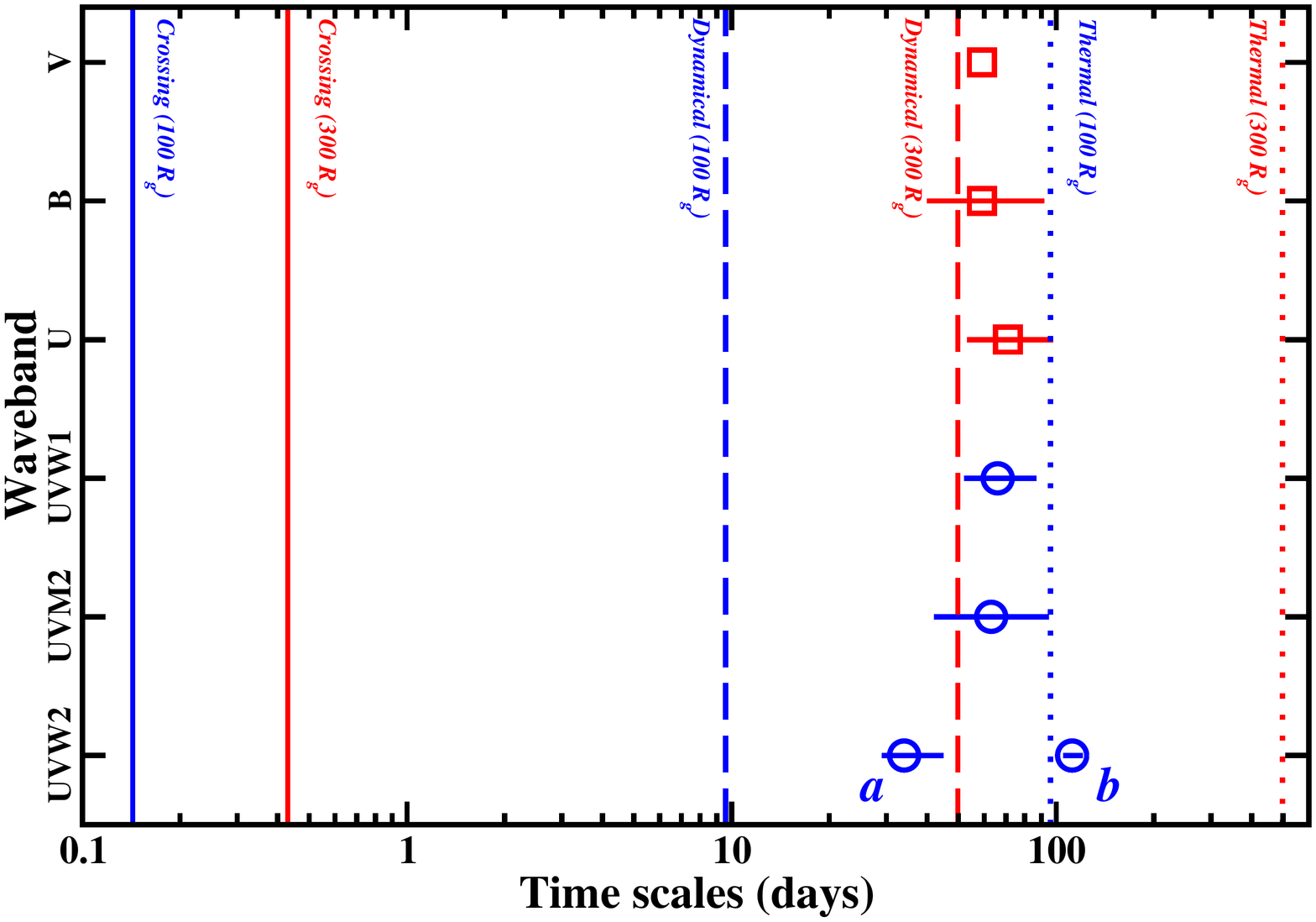}}
	\caption{Characteristic timescales determined from the structure function analysis are compared to standard accretion disc timescales at $100~R_g$ (blue lines) and $300~R_g$ (red lines). The distances correspond to where the UVW2 and B wavebands will be dominated by black body emission (i.e. UVW2 will be dominated by emission from $\sim100~R_g$ and B from $\sim300~R_g$). The measurements determined from UV light curves are shown with blue circles and those measurements with optical light curves are shown as red squares. Two measurements are shown for the UVW2 with (a) corresponding to  fitting the SF between $10-50$~days ($\tau\approx 34$~days) and (b) corresponds to fitting the data between $50-160$ days ($\tau\approx 112$~days) to demonstrate the possibility of two distinct slopes and characteristic timescales in the UVW2 SF.  A fit to the V-band data point correspond to the value of $\tau$ measured for the B-band. }
	\label{fig:chart}
\end{figure*}

The structure function slope ($\beta$) and characteristic timescale ($\tau_{char}$) are determined for the $\sim11$-year X-ray and UV-optical light curves of \mrk335.  For the UV-optical light curves, the SF slope is $\beta \approx 0.8$, roughly comparable to a PSD spectral index of $\alpha\approx 1.8$.  The values measured for \mrk335\ are within the range of $\beta$ found for a sample of Seyfert~1 galaxies (CP01) and comparable to the UV-optical PSD ($\alpha \approx 1-2$) often measured for Seyferts (e.g. Krolik \et 1991; Reichert \et 1994; McHardy \et 2004; Ar\'evalo \et 2008; Breedt \et 2010).  The similarity in the slopes of the UV and optical SF might suggest a common physical mechanism is at work.  

CP01 present only a V-band structure function for \mrk335.  A slope of $1.00\pm0.23$ and a characteristic timescale of $49^{+31}_{-23}$~days were measured from fitting the SF.  Unfortunately, the \swift\ V-band light curve is sparsely sampled and the amplitude of the variations in the band are small, rendering a poorly defined SF.  Independent measurements of $\beta$ and $\tau_{char}$ were not possible with the current data.  However, the CP01 SF measurements for \mrk335\ are consistent with the values measured here in the other optical bands (B and U).

Notably, the AGN Watch\footnote{http://www.astronomy.ohio-state.edu/$\sim$agnwatch/} monitoring campaign of \mrk335\ was conducted from the late-1980s to mid-1990s (Kassebaum \et 1997; Peterson \et 1998), prior to the NLS1 dropping to its current, long-lasting, low X-ray flux interval.  The similarities between the SFs calculated in this work with that of \mrk335\ in the mid-1990s, and with those of other Seyfert~1 galaxies, suggests the driving mechanism (i.e. accretion via a standard disc), has not fundamentally changed in \mrk335\ during this extended X-ray low-flux interval.  This may support a simple explanation like partial covering of the central region (e.g. Tanaka \et 2004). 

Under the assumption of a standard, Shakura-Sunyaev accretion disc (Shakura \& Sunyaev 1973), the temperature profile is 
\begin{equation}
         T\left(r\right)=\Bigg[\frac{3GM\dot{M}}{8\pi\sigma r^3}\Bigg(1 - \sqrt{\frac{r_{isco}}{r}}\Bigg)\Bigg]^{1/4}
\label{eq:tr}
\end{equation}
where $\sigma$ is the Stefan-Boltzmann constant.  At a given distance, the blackbody emission peaks at a wavelength
\begin{equation}
         \lambda\left(r\right)\approx\frac{hc}{3kT\left(r\right)}
\label{eq:lambda}
\end{equation}
Assuming the efficiency of converting mass to radiation is 0.1, the peak emission observed in a given bandpass will be dominated by thermal emission originating at a distance of 
\begin{equation}
\begin{split}
         \frac{r}{r_g}=\big(1.28\times10^{11}\big)\lambda^{4/3}\Bigg(\frac{M}{10^8\Msun}\Bigg)^{-1/3} & \Bigg(\frac{\dot{M}}{\dot{M}_{Edd}}\Bigg)^{1/3} \\
         & \Bigg(1 - \sqrt{\frac{r_{isco}}{r}}\Bigg)^{1/3}
         \end{split}
\label{eq:tr}
\end{equation}
where $\dot{M}$ is the accretion rate in units of the Eddington rate $\dot{M}_{Edd}$, $\lambda$ is in metres, and distance is in gravitational radii ($1r_g=GM/c^2$).

The peak, source-frame, wavelengths in the \swift\ $UVW2$ and $B$ wavebands are $1880\A$ and $4280\A$, respectively.  For \mrk335, a black hole mass of $M=2.6\times10^7\Msun$ (Grier \et 2012) and  $\dot{M} =0.2\dot{M}_{Edd}$ are assumed.   The emission in the UV and optical bands is dominated by blackbody emission that is originating at $\sim100r_g$ and $\sim300r_g$, respectively.

The timescales associated with standard accretion disc processes can be determined at these distances and compared to the timescales found in the SF and DCF analyses.
The light crossing time at a distance $r$ around a black hole of mass $M$ is
\begin{equation}
         t_{lc}=(5.5\times10^{-3})\bigg(\frac{M}{10^8\Msun}\bigg)\bigg(\frac{r}{r_g}\bigg) ~\rm{days}
\label{eq:lc}
\end{equation}
For \mrk335, $t_{lc}\approx0.14$~days in the UV region and $0.43$~days in the optical region.

The dynamical timescale is
\begin{equation}
         t_{dyn}=(3.7\times10^{-2})\bigg(\frac{M}{10^8\Msun}\bigg)\bigg(\frac{r}{r_g}\bigg)^{3/2}~\rm{days}
\label{eq:dyn}
\end{equation}
corresponding to 9.6~days (UV) and 50~days (optical) in \mrk335.

The thermal timescale, $t_{th} = t_{dyn}/\alpha_v$, where the viscosity parameter can be taken as $\alpha_v = 0.1$ in a standard accretion disc, will be $\sim96$~days (UV) and $\sim500$~days (optical).

Assuming a vertical-to-radial disc ratio of $h/r = 0.01$, the viscous timescales are $t_{vis} = \frac{t_{th}}{(h/r)^2} \approx 2600$~years and 13600 years in the UV and optical regions of \mrk335, respectively.  These timescales can be off by up to a factor of 10 depending on various assumptions, but generally the viscous timescales will be much longer than timescales considered in this work.

In Fig.~\ref{fig:chart}, the characteristic timescales ($\tau_{char}$) determined from each of the UV-optical structure functions is compared to the accretion disc timescales.  For all bands, the value of $\tau_{char}$ is similar (Table~\ref{tab:sf}), though the UV emission originates a third of the distance closer to the black hole than the optical emission.  The optical timescales agree very well with the orbital timescales at $300r_g$ in the standard disc.  For the UV wavebands, the 
characteristic timescales are most likely comparable to the thermal timescales at $\sim100r_g$ in a standard accretion disc.  Indeed, if uncertainties in the black hole mass were considered, along with the range of radii that are examined in each waveband, then the agreement would be even better.  

The $UVW2$ light curve of \mrk335\ is of substantial high-quality that the SF is rather detailed. The SF appears to possess two breaks in the power spectrum.  At short-timescales, the SF seems to match up well with the other bands in terms of its slope, but breaks at a shorter characteristic timescale of $\sim34$~days.  The function steepens at longer timescales before again breaking at $\sim112$~days.  Only the longer timescale is comparable to the thermal timescales in the UV region of the disc.  The shorter break is not comparable to any obvious accretion disc timescale and only consistent with dynamical timescales at $\sim230\rg$ or thermal timescales if $\alpha_v$ were $\sim 0.3$.   Both are notably different than the $\tau_{char}$ derived from the other UV bands.  

The slope of the SF between $10-50$~days ($\beta=0.86^{+0.14}_{-0.13}$) is flatter than the slope between $50-160$~days ($\beta=1.25^{+0.20}_{-0.18}$).  Though both slopes are consistent with the average slope measured in all \uvot\ bands, they are different from each other.  This may be indicating that the breaks are unique and that the power spectra are arising from independent processes.  It could be the second break exists in all wavebands, but not significantly detected because of limited data or data quality.

CP01 noticed multiple breaks in the optical structure function of NGC~5506 and suggested they may be attributed to starburst activity. Aretxaga \et (1997) generated models of the expected variability in a young stellar cluster with compact supernova remnants and predicted characteristic timescales between $85-280$~days that may be consistent with observations.  It is unclear if a similar interpretation can be adopted for the UV variability in \mrk335\ given the low star formation rate implied by its weak PAH emission (e.g. Sani \et 2009). 

Alternatively, the existence of a compact or abort jet (e.g. Ghisellini \et 2004) in \mrk335\ as previously suggested (Gallo \et 2013, 2015; Wilkins \et 2015) may demand other timescales are considered.  
Though \mrk335\ is radio-quiet and not known to have a jet, X-ray flaring during Year 8 (Wilkins \et 2015; see below) was attributed to collimated, relativistic outflow that could also generate UV emission.  Indeed, the DCF between UVW2 and X-rays was strongest in Year 8.  The timescales detected here may be too large even within a jet that cools as it propagates (i.e. with the UV emitted upstream from the X-ray emission point).

One could examine the UVW2  structure function in individual years to determine any variable behaviour in the function itself.  This would be challenging despite how good these data are.  \mrk335\ is observed for $\sim240$~days of each year.  The SF is robust within $\sim0.3T$ (CP01), meaning the slope can be reliably measured up to $\sim70$~days. Each year would allow investigation of only the short timescale SF that is apparent in the 11-year data set.  Moreover, the SF data quality (i.e. resolution and scatter) would be compromised as fewer light curve pairs ($i, j$) would contribute to each SF bin.

The X-ray structure function has a shallow power law slope consistent with 0 (flat)  between $\sim2-50$~days.  There is no observable break, which may be indicative of a turnover occurring at much higher frequencies (Ar\'evalo \et 2008).  Ar\'evalo \et measured a break in the power spectrum at $\sim1.7\times 10^{-4}$~Hz ($\tau \approx 6000\s$) in the 2006 \xmm\ observation of \mrk335\ when the AGN was in a bright flux state.  At lower frequencies the slope was fixed at $\alpha = 1$ though allowing it to vary did not significantly change the value.  The timescale is in relatively good agreement with black hole mass measurements for \mrk335.

The SF slope measured here is comparable to a PSD slope of $\sim 1$, consistent with the measurements of  Ar\'evalo \et (2008).  Assuming the SF power law is comparable to the extrapolation of the flat low-frequency, power law spectrum measured by Ar\'evalo \et (2008), there is  no strong indication of physical changes in the processes causing the X-ray emission.

The search for correlations and lags in the 11-year light curves of \mrk335\ did not yield highly significant results, and most correlations were at the $\sim95$~per cent confidence level.  The lack of strong correlations was consistent with previous findings in Grupe \et (2012) using a shorter light curve.  However, there is evidence the strength of the correlations between the X-rays and UVW2 changes across the light curve.  For example, in Year 8 of the campaign, the UVW2 and the X-rays were well-correlated at $\sim99.9$ per cent confidence level.  There was also a slight shift in the peak of all the DCFs indicating the X-ray variations lead the fluctuations in other lower energies wavebands. In all cases, the shift was equivalent to one resolution element in the DCF (i.e. $5\pm5$~days) and formally consistent with zero lag.   However, monitoring with higher cadence could reveal significant correlations.

During Year-8 (August 2014), \mrk335\ exhibited a giant (factor of 10) X-ray flare (Grupe \et 2014) that triggered a \nustar\ target-of-opportunity observation.  The X-ray flare was interpreted as vertically collimated outflow moving away from the disc at mildly relativistic speeds (Wilkins \et 2015).  It was noted that the UVW2 emission peaked and diminished along with the X-ray flare, though the flaring in the UV was only at the $\sim25$~per cent level.  Since the number of corona photons striking the disc compared to the total number of corona photons emitted (i.e. the reflection fraction) did not change during the X-ray flare, Wilkins \et (2015) disfavour X-ray reprocessing as the cause of the UV brightening.  Instead, the UV emission could be generated in a jet-like structure through synchrotron self-Compton processes (e.g. Ghisellini \et 1985).  

The data set in Year 2 is important because all the \uvot\ wavebands are equally well sampled in that epoch and correlations between all the wavebands could be investigated.  As with the 11 year average DCFs, no correlations were found at greater than the $95$~per cent confidence level between the X-ray and any of the UV-optical wavebands.  However, mores significant correlations at  $>95$ per cent confidence level were found between the UVW2 light curve and other UV-optical light curves.  Again, a slight offset of one resolution element, corresponding to a lag of $0-5$~days is seen with the UVW2 light curve leading the others. The temporal resolution is not sufficient to investigate lags shorter than 5~days, but the timescales are what may be expected from X-ray reprocessing  in the standard accretion disc that follows a $\tau_{\rm{lag}} \propto \lambda^{4/3}$ relation (e.g. McHardy \et 2014).  Such lags would be short for \mrk335, only part of a day, but many reverberation studies also show the amplitude of the lags  imply discs that are larger than predicted by the standard model (e.g. McHardy \et 2014; Cackett \et 2017).


\section{Conclusions } 
\label{sect:conc}

\swift\ monitoring of \mrk335\ since 2007 while the NLS1 has been in an extended low-X-ray-flux state are presented.  Analysis of the UV-optical and X-ray structure functions (power spectra) suggest the AGN is consistent with previous high-flux observations of it, and comparable with other Seyfert~1 galaxies.  Variability in the UV and optical emission can be attributed to thermal and dynamic processes in the accretion disc, respectively.  The current X-ray low flux state of \mrk335\ does not appear to be driven by changes in the accretion disc structure or behaviour.  The changes originate in the X-ray emitting region and could be attributed to physical changes in the corona or absorption. Interestingly, the UVW2 ($\sim1800\A$) structure function possesses two breaks and two different slopes indicating multiple physical processes are at work.  We speculate that the presence of a unresolved, compact (perhaps aborted) jet may be at work in \mrk335\ and responsible for the appearance of the structure function.  

The search for correlations and lags between different energy bands revealed interesting behaviour that should be followed up with more frequent monitoring of \mrk335\ with shorter cadence.  Correlations are not significant beyond the $\sim95$ per cent confidence level when considering the 11-year light curves, but more significant behaviour is present when considering segments of the light curves.  A correlation between the X-ray and UVW2 in 2014 (Year-8) may be predominately caused by an giant X-ray flare that was interpreted as jet-like emission.  In 2008 (Year-2), more significant correlations ($>95$ per cent level) and possible lags between the UVW2 emission and other UV-optical wavebands may be consistent with reprocessing of X-ray/UV emission in the accretion disc.

\section*{Acknowledgments}
We thank the \swift\ team for approving our various ToO requests to monitor \mrk335\ over the years.
LCG and DMB are grateful to Dr. Tina Harriott for discussion and support.  Many thanks to William Alston and the referee for
providing suggestions that improved the manuscript.


\bsp
\label{lastpage}
\end{document}